\documentstyle{mn}

\input{epsf}

\def\comptel{{\it COMPTEL~}}
\def\cgro{{\it CGRO~}}
\def\egret{{\it EGRET~}}
\def\rosat{{\it ROSAT~}}

\def\glast{{\it GLAST~}}
\def\magic{{\it MAGIC~}}

\def\ergscm2{~{\rm erg}~{\rm s}^{-1}~{\rm cm}^{-2} }

\def\MeV{~\rm{MeV}}
\def\GeV{~\rm{GeV}}
\def\TeV{~\rm{TeV}}
\def\cm{~\rm{cm}}
\def\K{~\rm{K}}
\def\G{~\rm{G}}
\def\Lx{$L_x$}
\def\Lg{$L_\gamma$}
\def\edot{$L_{\rm sd}$}
\def\s{\rm{s}}
\def\ms{\rm{ms}}
\def\lambdabar{\mathrel{\lower 1pt\hbox{$\mathchar'26$}\mkern-9mu
        \hbox{$\lambda$}}}

\newbox\grsign \setbox\grsign=\hbox{$>$} \newdimen\grdimen \grdimen=\ht\grsign
\newbox\simlessbox \newbox\simgreatbox \newbox\simpropbox
\setbox\simgreatbox=\hbox{\raise.5ex\hbox{$>$}\llap
     {\lower.5ex\hbox{$\sim$}}}\ht1=\grdimen\dp1=0pt
\setbox\simlessbox=\hbox{\raise.5ex\hbox{$<$}\llap
     {\lower.5ex\hbox{$\sim$}}}\ht2=\grdimen\dp2=0pt
\setbox\simpropbox=\hbox{\raise.5ex\hbox{$\propto$}\llap
     {\lower.5ex\hbox{$\sim$}}}\ht2=\grdimen\dp2=0pt
\def\simgreat{\mathrel{\copy\simgreatbox}}
\def\simless{\mathrel{\copy\simlessbox}}

\begin{document}

\label{firstpage}

\title{Spectral features in gamma-rays expected from millisecond pulsars}

\author[Bulik,  Rudak and  Dyks] {T. Bulik$^1$,
B. Rudak$^{2,3}$ and J. Dyks$^2$\\
 $^1$ Nicolaus
Copernicus Astronomical Center, Bartycka 18, 00716 Warsaw, Poland\\ 
$^2$ Nicolaus
Copernicus Astronomical Center, Rabia{\'n}ska 8, 87100 Toru{\'n}, Poland\\
$^3$ TCfA of Nicolaus Copernicus  University, Gagarina 11, 87100 Toru{\'n}, Poland}
\maketitle 

\begin{abstract} 

In the advent of next generation gamma-ray missions, we present
general properties of spectral features of high-energy emission
above $1\MeV$ expected for a class of millisecond, low magnetic
field ($\sim 10^9\G$) pulsars.  We extend polar-cap model
calculations of Rudak \& Dyks (1999) by including inverse Compton
scattering events in ambient field of thermal X-ray photons and
by allowing for two models of particle acceleration.  In the
range between $1\MeV$ and a few hundred GeV the main spectral
component is due to curvature radiation of primary particles.
Synchrotron component due to secondary pairs becomes dominant
only below $1\MeV$.  The slope of the curvature radiation
spectrum in the energy range from $100\MeV$ to $10\GeV$ strongly
depends on the model of longitudinal acceleration, whereas below
$\sim 100\MeV$ all slopes converge to a unique value of $4/3$ (in
a $\nu {\cal F}_\nu$ convention).  The thermal soft X-ray
photons, which come either from the polar cap or from the
surface, are Compton upscattered to a domain of the VHE and form
a separate spectral component peaking at $\sim 1\TeV$. 
We discuss observability of millisecond pulsars by future
high energy instruments and present two rankings
relevant for GLAST and MAGIC. We
point to the pulsar J0437-4715 as a promising candidate for
observations. 

\end{abstract}


\date{}

\section{Introduction}

Millisecond pulsars are thought to be
uninteresting  as targets for gamma-ray experiments. This
opinion partly draws from an argument that typical values
of $B$ (or rather their dipolar components) inferred from
$P$ and $\dot P$ are in the range of $10^8$ to $10^9\G$, in
contrast with the range of $10^{11}$ to $10^{13}\G$ for
classical pulsars. If gamma-ray radiation is to be a
manifestation of magnetospheric activity, it seems quite
natural to think of millisecond pulsars in terms of $B$ as
of a scaled-down version of classical pulsars, which by
themselves are weak gamma-ray sources; only seven
of them have been firmly detected so far
\cite{1997comp.symp...39T}. Unsuccesful observational
campaigns in the past, mainly by \cgro instruments e.g.
\egret - \cite{1995ApJ...447..807F}, confirmed this line
of thinking.  J0437-4715, with one of the highest spin-down
fluxes among all pulsars, thus a very promising target,
wasn't even included in the \comptel priority list
\cite{1995A&A...304..258C}.

However, the `scaling-down' argument becomes unjustified by the
results of X-ray observations.  Within a sample of some 30
pulsars detected with \rosat and other satellites there are 9
millisecond objects \cite{1999A&A...341..803B} with apparent
X-ray properties similar to classical pulsars.  In particular,
the millisecond pulsars obey the same astonishing relation of
Becker~\&~Tr\"umper \shortcite{1997A&A...326..682B} between the
spin-down luminosity \edot~ and the soft X-ray luminosity \Lx~
that classical pulsars do.  In most cases the X-ray spectra are
presumed to be non-thermal in character, but whether they
represent low-energy tails of putative gamma-ray emission is not
known.

Rather few papers have been dedicated to the question of
gamma-rays in the context of millisecond pulsars on theoretical
side.  Wei et~al.  \shortcite{1996ApJ...468..207W} presented
broad-band spectra of both pulsed and unpulsed emission expected
in their version of outer-gap model.  Sturner \& Dermer
\shortcite{1994A&A...281L.101S} offered predictions for gamma-ray
luminosity \Lg~ in some popular models, but that had been done
just by scaling-down the values of $B$ in simple analytical
formulae for \Lg~ derived for classical pulsars.  In the
framework of polar-cap scenarios, Rudak \& Dyks
\shortcite{1998MNRAS.295..337R} proposed a modification of the
model by Daugherty \& Harding \shortcite{1982ApJ...252..337D}.
This modification incorporates a contribution of
electron-positron pairs to \Lg, which becomes a non-monotonic
function of $B$ (see Figures 7 and 8 of
\cite{1998AcA....48..355D}).  One of the consequences of this
modification is that expected values of \Lg~ for millisecond
pulsars, including J0437-4715, are typically about one or two
orders of magnitude lower than for classical pulsars.
Quantitatively similar estimate, though for different reason,
comes from a model of Dermer \& Sturner
\shortcite{1994ApJ...420L..75D}.

However, if the estimates above are correct then some millisecond
pulsars, including J0437-4715, should be detected with
next-generation experiments, providing thus a testing ground for
magnetospheric models relevant for millisecond pulsars.  The
successor of \egret~ - \glast - with a possible launch in the
year 2005, is expected to have sensitivity above $100\MeV$ about 30
times higher then \egret, and it will reach high-energy limit at
$300\GeV$ - closing thus for the first time the energy gap
between the VHE (very high energy) domain accessible with
ground-based Cherenkov techniques and the HE (high energy) domain
of satellite experiments \cite{1999astro-ph}.  Moreover, since we
expect the millisecond pulsars to radiate the electromagnetic
energy mostly in the range between $10\GeV$ and $100\GeV$, this
class of objects should be of interest for the upcoming IAC
telescope {\it \magic} \cite{1998..Magic}.

Our aim is to present major spectral features in the energy range
above $\sim 10\MeV$ of pulsed gamma-ray emission which arises due
to both ultrarelativistic primary particles (electrons) and
cascades of secondary particles above polar caps of millisecond
pulsars with dipolar magnetic fields of $\sim 10^9\G$.  In
particular, we allow for different structures of the accelerating
potential, and we show how this affects the slope of the
predicted energy spectrum in the range of $100\MeV$ to $100\GeV$.
Also, Compton upscattering of ambient X-ray photons off
ultrarelativistic electrons is taken into account.  In Section 2
we present details of modelling the processes responsible for the
formation of gamma-rays in millisecond pulsars.  Section 3
describes spectral properties of two distinct components, due to
curvature radiation (CR) and Compton scattering (ICS).
Conclusions with a discussion of J0437-4715 come in Section 4.

\begin{figure*}
{\centering \leavevmode
\epsfxsize=.45\textwidth \epsfbox{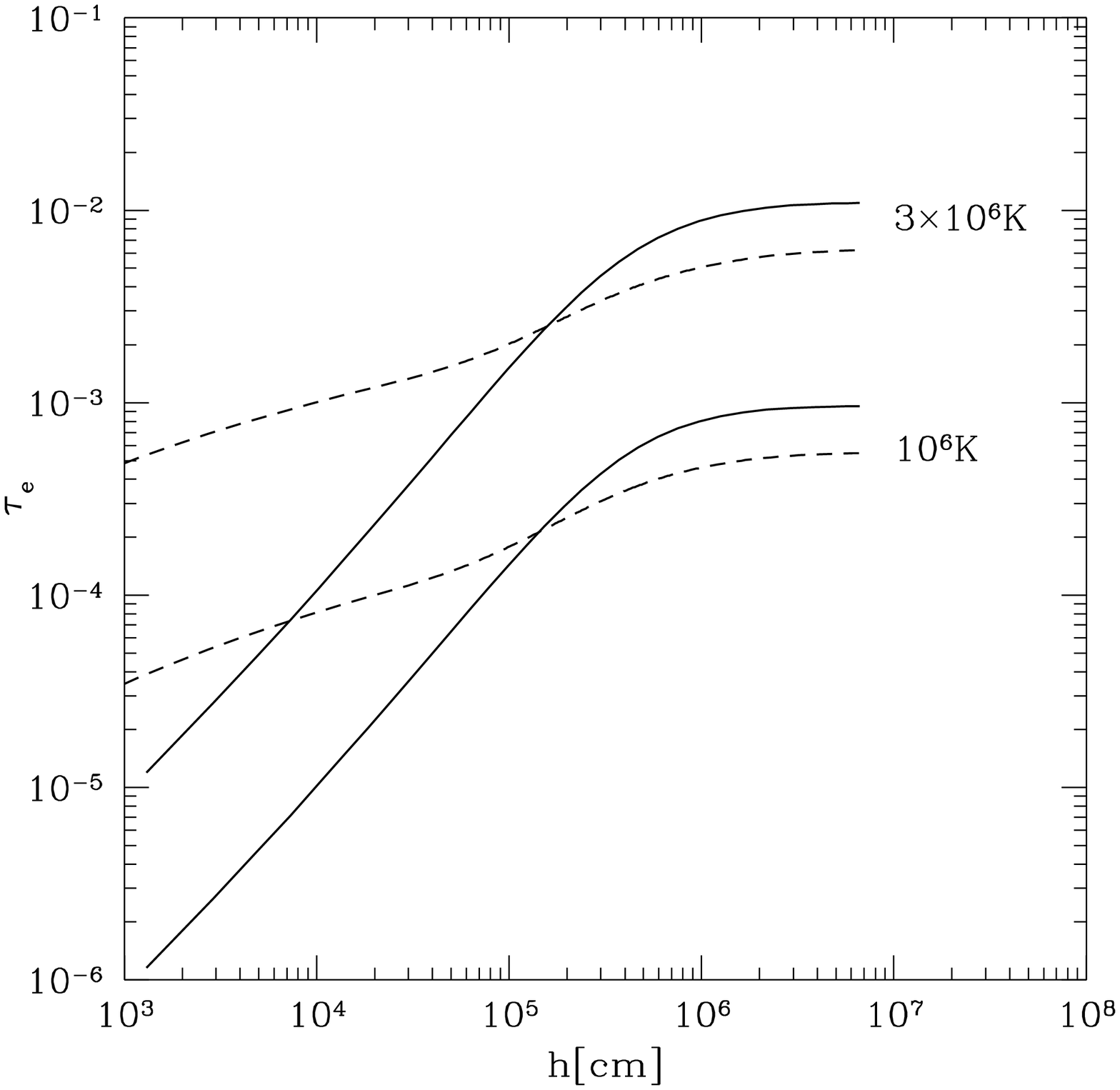} \hfil
\epsfxsize=.45\textwidth \epsfbox{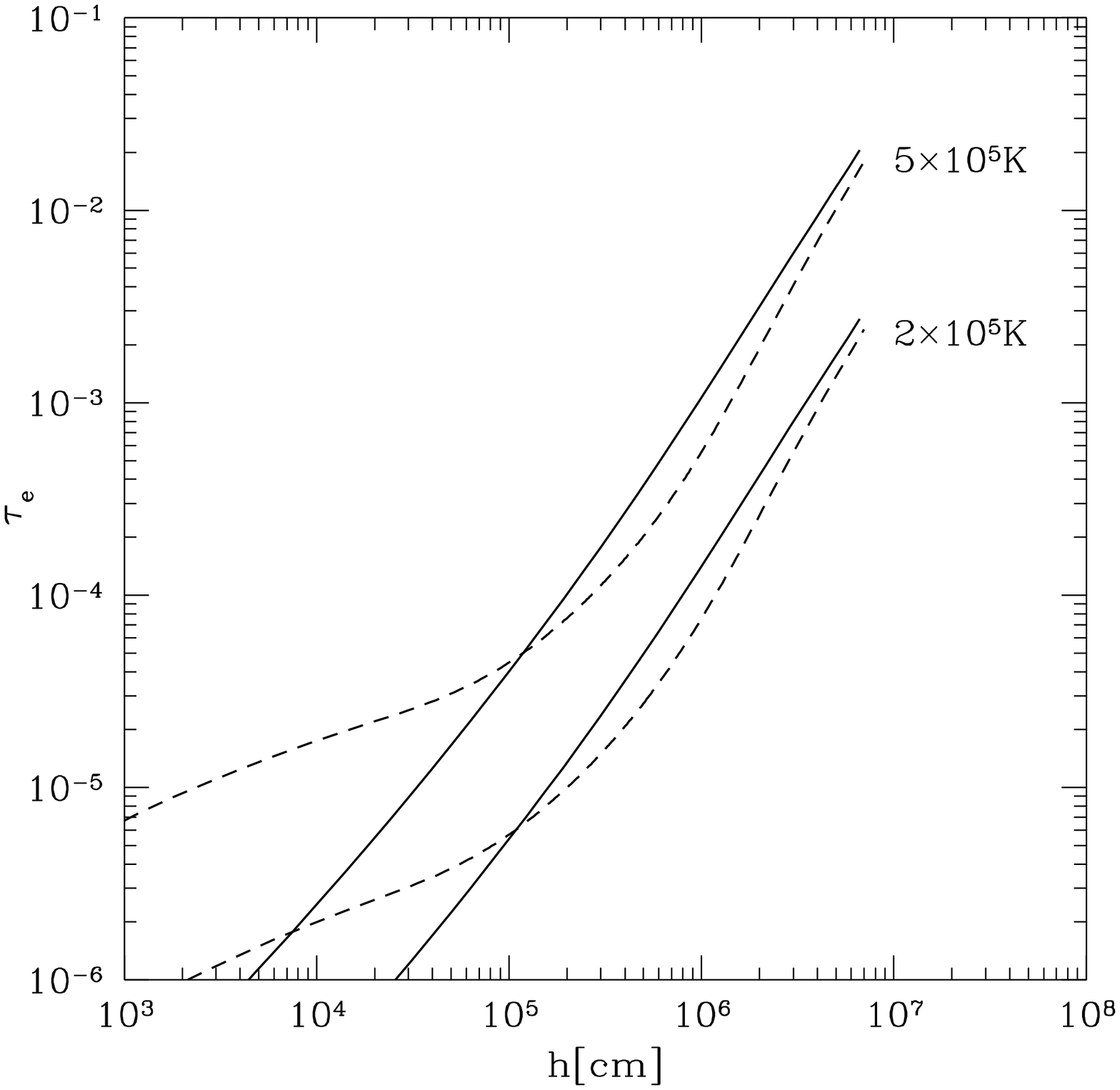}}

\caption{Cumulative optical depth $\tau_e$ for non-magnetic scattering of an electron
injected at the surface level ($h = 0$) and crossing a blackbody 
radiation field. The left panel is for the radiation fields due to
hot polar cap ($T_{\rm cap} = 10^6\K$ and $3 \times 10^6\K$).
Solid lines are for instant acceleration with electron
energy $E_{\rm init}= 10^7\MeV$. Dashed lines are for continuous
acceleration in electric field $\cal E$ (eq.\ref{efield}).
The right panel is for the radiation fields due to hot stellar surface
($T_{\rm surf} = 2 \times 10^5\K$ and $5 \times 10^5\K$).
The pulsar period is $P = 3\times 10^{-3}\s$ and we used $10^5$ 
electrons in the simulation.
}
\label{fig1}
\end{figure*}

\section{The model of high-energy radiation}

We use a polar cap model, with most ingredients as postulated by
Daugherty \& Harding (1982) in context of classical pulsars.
High-energy emission is a superposition of curvature radiation
(CR) of ultrarelativistic primary electrons, and synchrotron
radiation (SR) of secondary particles (e$^\pm$ pairs) created via
magnetic absorption of photons.  Additional features include:  1)
two models of electron acceleration, 2) inverse Compton
scattering (ICS) of thermal X-ray photons on electrons.  The
first feature is important for the history of cooling of primary
electrons and consequently it affects the slope of gamma-ray
spectrum above $100\MeV$.  The second one (ICS) is not important
for overall cooling of electrons, but the resulting TeV component
in the gamma-ray spectra is of potential interest.  A field of
ambient photons necessary for the ICS is likely to be present
within the magnetosphere of some millisecond pulsars, according
to some interpretations of available X-ray and EUV observations
(e.g.  Zavlin \& Pavlov 1998, Edelstein et al.  1995).  We list
the details of the model in the following subsections.

\subsection{Geometry}

We assume that the geometry of the magnetic field of a neutron
star is well described by a static, axisymmetric dipole.  In
polar coordinates its field lines satisfy
\begin{equation} 
r \sin^{-2}\theta = R_0,
\end{equation}
where the dipole constant $R_0$ defines a set of field lines
which differ by azimuthal angle only.  Polar caps are defined as
(two) regions on the neutron star surface crossed by all lines
for which the condition $R_0 \geq R_{\rm lc}$ is satisfied, where
$R_{\rm lc}=c P/2\pi$ is a light-cylinder radius for a spin
period $P$.  The polar cap radius is then $r_{\rm pc} = R_{\rm
ns}\times(R_{\rm ns}/R_{\rm lc})^{1/2} $.  where $R_{\rm ns}$ is
the radius of the neutron star for which we assume the canonical
value of $10^6\,$cm.  Beam particles (primary electrons) are
ejected from the outer rim of the polar cap and move along the
magnetic field lines.

When modelling an extended source of soft photons, which are to
be targets for electrons in Compton scatterings, we consider two
cases of its geometry.  In the first case, the soft photons come
from the hot polar cap with a uniform temperature $T_{\rm
pc}\simgreat 10^6\K$.  In the second case the soft photons come
from the entire surface of the neutron star, however now the
temperature is lower:  $T_{\rm surf}\simgreat 10^5\K$.

\subsection{Electron acceleration}

Primary electrons are injected along magnetic field lines into
the magnetosphere from the outer rim of the polar cap.  The
structure of longitudinal electric field responsible for their
acceleration remains unknown.  However, the electric field must
be strong enough to ensure electron positron pair creation.  The
presence of pairs is a necessery condition for a magnetized
neutron star to be a radio pulsar.  Thus the requirement of
creating pairs in the magnetosphere is the primary condition that
the model has to satisfy.

We use two simple models for particle acceleration.  The first
one (denoted model A) assumes instant acceleration to
ultrarelativistic energy $E_{e}^{\rm init}$ at the moment of electron
injection on the surface of the neutron star.  The values of
$E_{e}^{\rm init}$ are set to be just above the threshold value for
creation of secondary pairs $E_{e}^{\rm min}$ which in turn is a
function of magnetic field strength at the polar cap $B_{\rm pc}$
and the spin period $P$ (for details see Rudak \& Dyks
1998, and Dyks 1998).  For a pulsar with $B_{\rm pc} =10^9\G$ and
$P = 3\ms$, the initial energy is $E_{e}^{\rm init} = 1.07\times
10^7\MeV$.  For J0437-4715 we took $B_{\rm pc} =7.4\times 10^8\G,
\, P = 5.75\ms$, and $E_{e}^{\rm init} = 1.54\times 10^7\MeV$, where
we used the magnetic field derived from $\dot P$ kinematically corrected
by Camilo et~al.~(1994).

For comparison we consider another model (model B) of particle
acceleration, in which the longitudinal electric field $\cal E$
is set at a constant level with an exponential cutoff:
\begin{equation}
{\cal E} = {V_0\over r_{\rm pc}} \exp\left(-{h\over r_{\rm pc}}\right), 
\label{efield}
\end{equation}
where the characteristic scale-height of the electric field is
assumed to be equal to the radius of the polar cap $r_{\rm pc}$.
For a pulsar with $B_{\rm pc} =10^9\G$ and $P = 3\ms$ the value
of $\cal E$ at the polar cap surface was set at the level of
$V_0/r_{\rm pc} \approx 10^{9}$V~cm$^{-1}$ to ensure a total
yield of secondary pairs per primary electron similar to that in
model (A).  The initial value of the electron energy $E_{e}^{\rm
init} = 10\MeV$ was assumed.

As electrons accelerate they move along the
magnetic field lines and are slowed down by the curvature
radiation reaction. We calculate the electron energy budget
as a function of the distance  travelled in the
magnetopshere. Only a small fraction (less than one  percent,
see Figure~\ref{fig1}) of the electrons will scatter on
thermal photons originating on the polar cap or on the stellar
surface. Thus scattering is not an important electron
deceleration process.

\begin{figure}

{\centering \leavevmode
\epsfxsize=\columnwidth \epsfbox{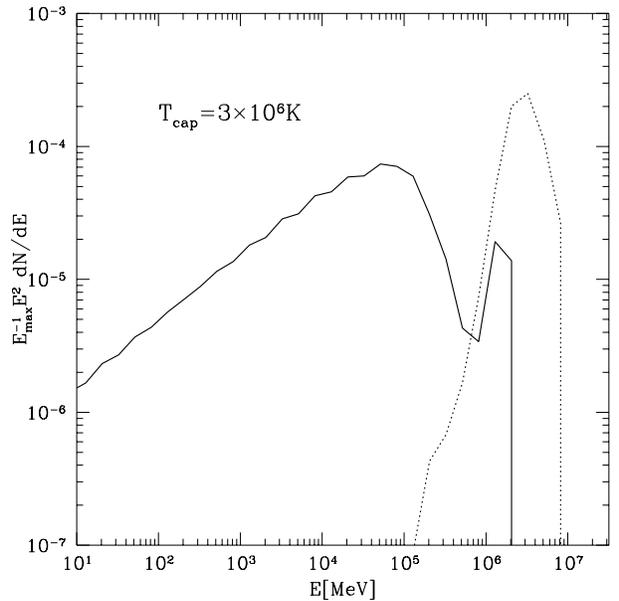}}
\caption 
{The dotted line shows the intrinsic spectrum of the
inverse Compton photons i.e.  with transfer effects neglected.
The solid line shows the  spectrum of magnetically
reprocessed photons    as seen by an external  observer.
Here we assumed:  thermal emission from a polar cap with
the temperature $T_{\rm cap}=3\times 10^6\K$, surface
magnetic field $B=10^9\G$, pulsar period $P=3\times
10^{-3}\s$, and instantaneous electron acceleration.}

\label{fig2}
\end{figure}

\begin{figure*}
{\centering \leavevmode
\epsfxsize=.45\textwidth \epsfbox{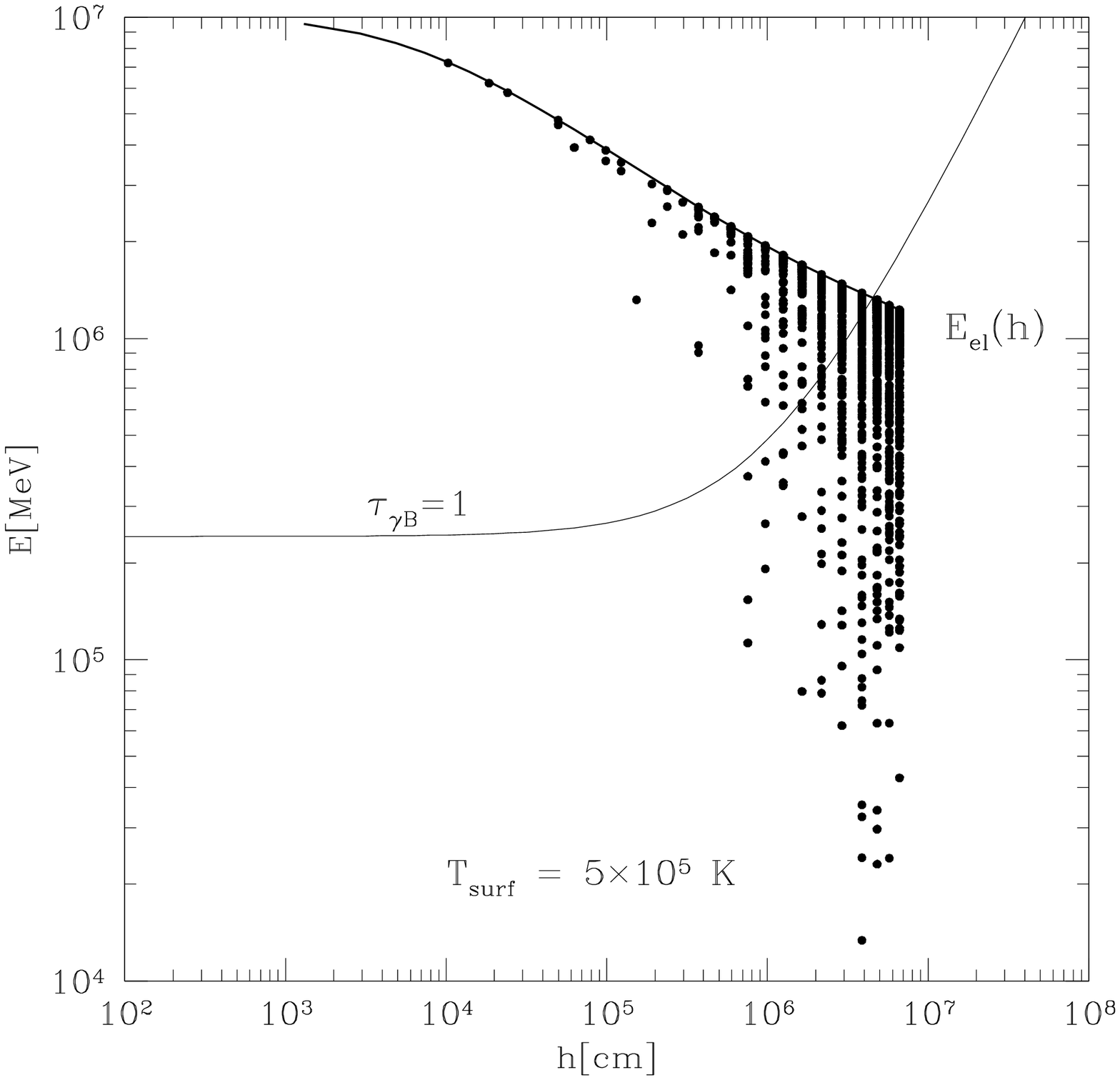} \hfil
\epsfxsize=.45\textwidth \epsfbox{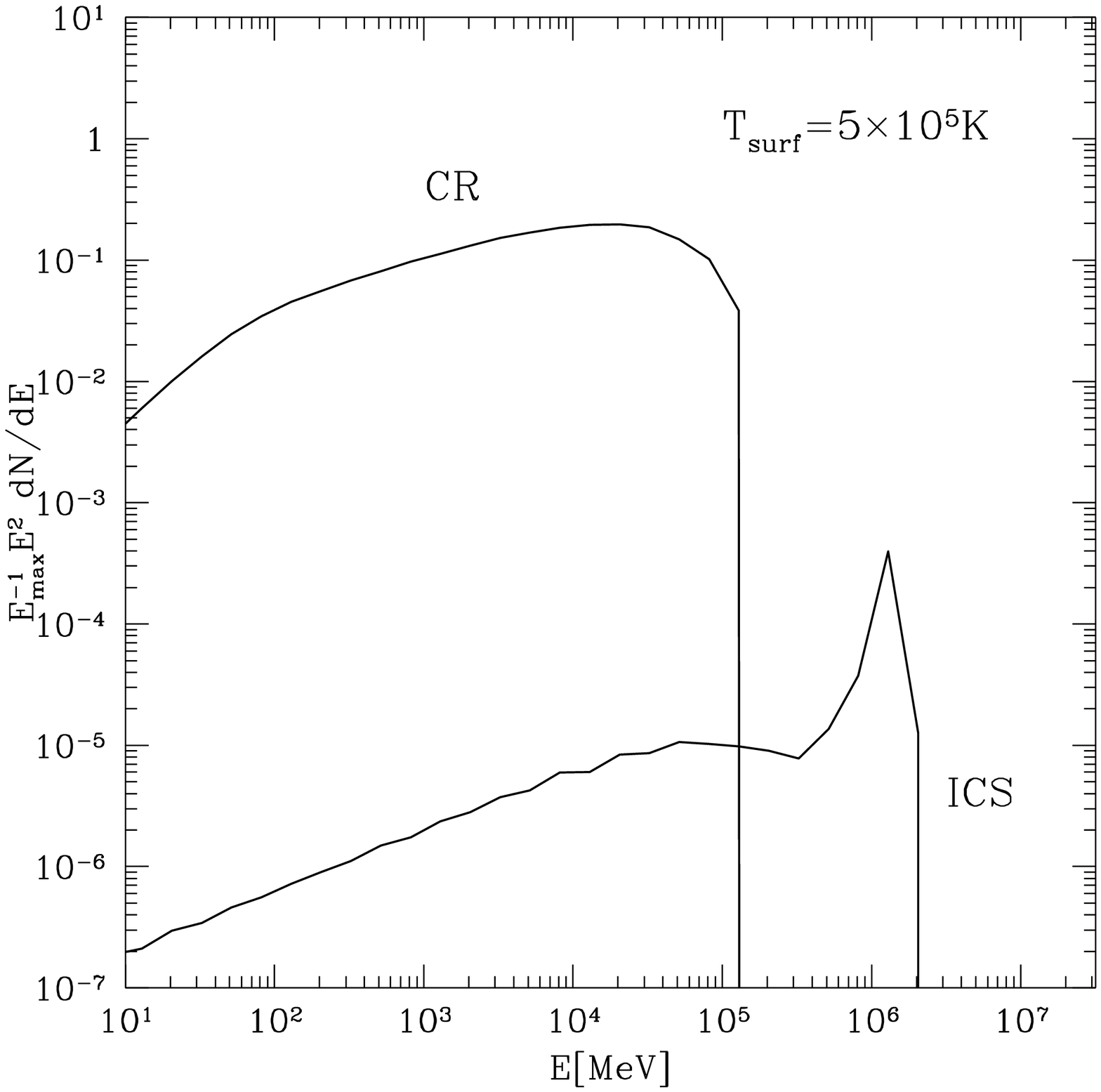}}
{\centering \leavevmode
\epsfxsize=.45\textwidth \epsfbox{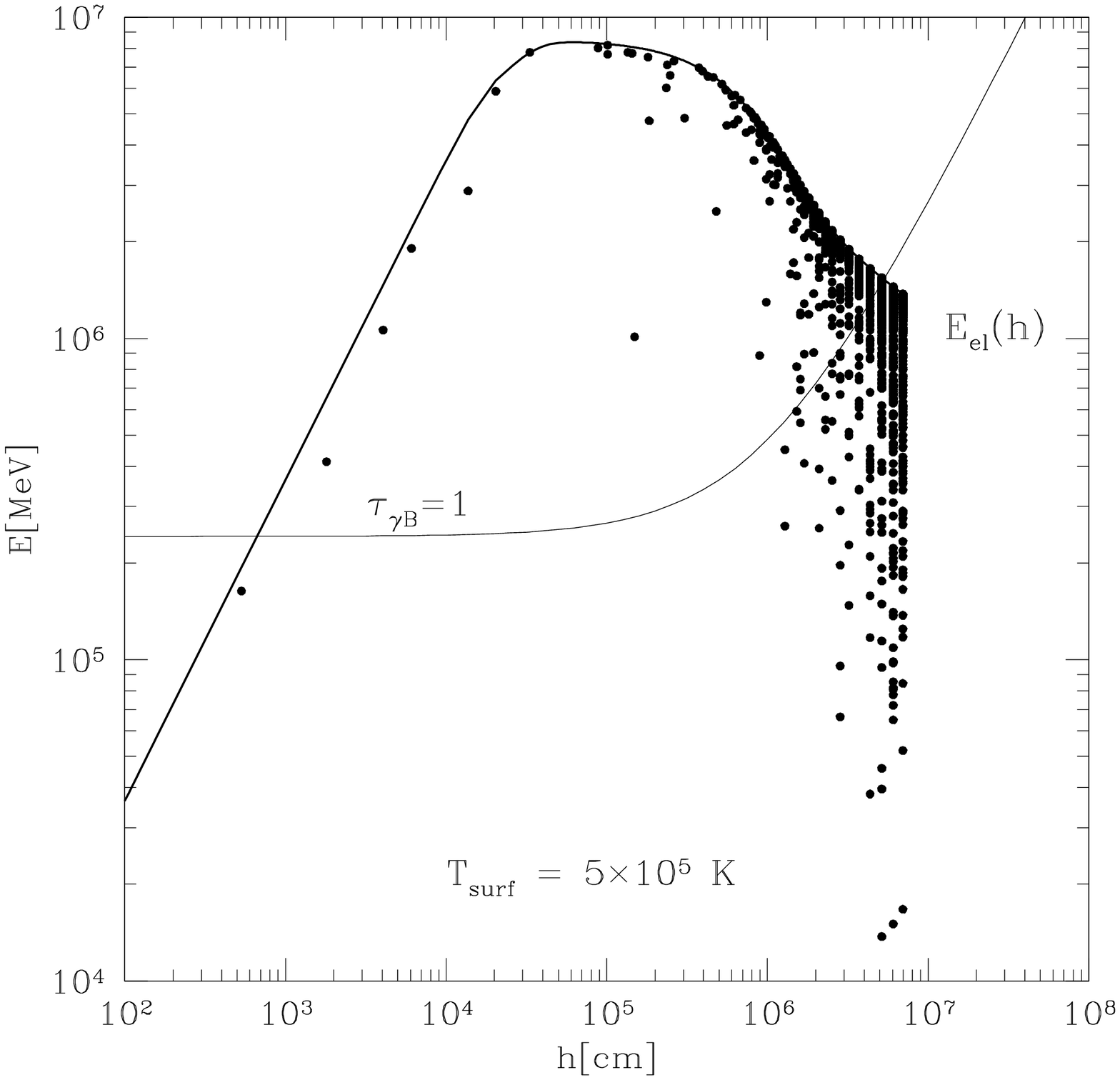} \hfil
\epsfxsize=.45\textwidth \epsfbox{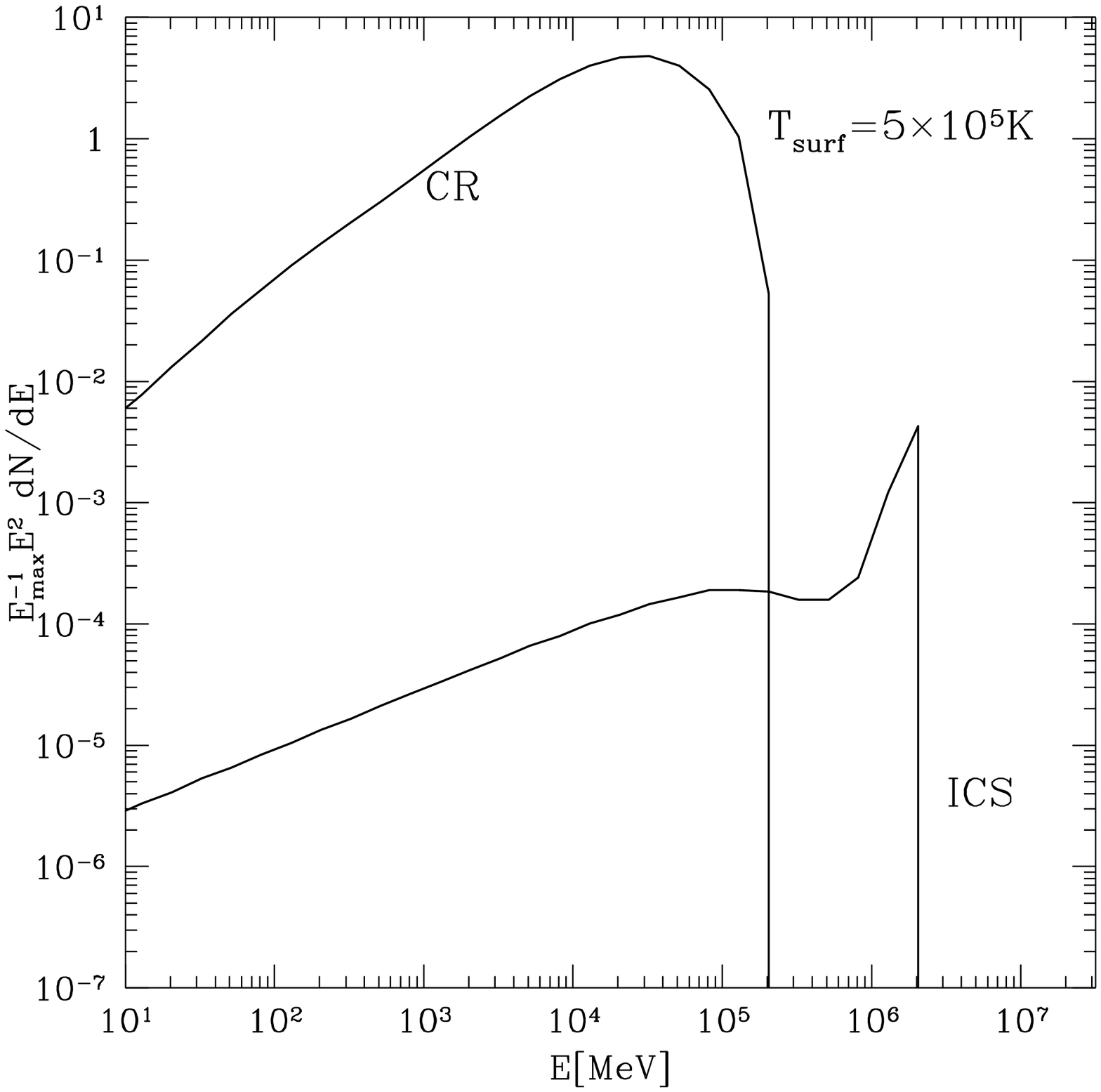}}
\caption{ 
The thick solid lines in the left panels show the dependence of
the   mean energy of unscattered primary electrons as a function of the height. The top left
panel corresponds to the case of instantaneous  acceleration (model A),
while the bottom left panel corresponds to case of accleration
described by eq.~(\ref{efield}) (model B). The dots show the energies  of
the Compton upscattered photons and height where the scatterings
took place (in these simulations we propagated  $10^5$ 
primary electrons in the magnetosphere). The thin solid line shows  the optical depth to
magnetic absorption equal to unity, according to equation~(\ref{tauone}). The right panels show the
energy spectra formed for model A (top
right panel) and model B
(bottom right panel). The soft photons originate on the surface
of the neutron star with the temperature $T_{\rm surf}=5\times 10^5\K$. 
The CR and the ICS  spectral components are due to a
single primary electron injected  at the outer rim of the polar
cap of the standard millisecond pulsar. We normalise the spectra
with $E_{e}^{\rm max}$ - maximum  energy of the electron, mediated
between acceleration and cooling rates.  For the top left 
panel, $E_{e}^{\rm max}$ is equal to the initial energy $E_{e}^{\rm init} =
1.07\times 10^7\MeV$. Pulsar parameters are $B_{\rm pc} = 10^9\G$, and $P =
3 \ms$ and $E_{e}^{\rm init}=1.07\times 10^7\,$MeV for model A.} 
\label{fig3} 
\end{figure*}

\subsection{Curvature and synchroton processes}

We follow the model described by Rudak \& Dyks
\shortcite{1999MNRAS.303..477R} who calculated broad band energy
spectra of high-energy emission for two distinct pulsar groups:
low-magnetic field pulsars (`millisecond pulsars') and classical
pulsars.

The dominant cooling process for primary electrons is the
curvature radiation:  
\begin{equation} 
(\dot \gamma_{e})_{\rm cr} =
(2/3) \gamma_{e}^4 \rho^{-2} c r_0 , \label{curv}
 \end{equation}
where $\gamma_{e}$ is the electron Lorentz factor,
$\rho$ is the curvature radius, and $r_0$ is the electron
radius.  When considering secondary electrons (pairs) one has to
consider also the synchrotron radiation.  The rate of SR cooling
is
\begin{equation}
(\dot \gamma_{e})_{\rm sr} =  - {2 \over 3} \,
{r_0^2 \over m_{\rm e} c}\, B^2 \sin ^2\psi \,\gamma_{e}^2\, ,
\label{SR0}
\end{equation}
where $\psi$ is the pitch angle of the pairs.
The photon spectra due to curvature and synchrotron processes are
calculated following Daugherty \& Harding
\shortcite{1982ApJ...252..337D}.  A detailed dicussion of the
role of these processes in the formation of gamma-ray spectra of
pulsars is given in Rudak \& Dyks
\shortcite{1999MNRAS.303..477R}.

\subsection{Scattering}

Our numerical treatment of scattering process follows Daugherty
\& Harding (1989).  As the electrons travel up in the
magnetosphere they may scatter off thermal photons originating
from the stellar surface or the polar cap.  For millisecond
pulsars the magnetic field at the surface is typically $10^9\G$
or less, i.e.  much weaker then the critical field $B_{crit} =
4.414\times 10^{13}\G$.  Thus the scattering process will be well
described by the Klein-Nishina relativistic nonmagnetic cross
section:

\begin{eqnarray}
\sigma_{KN} &=&
{3\sigma_T\over 4x} \left[{x^2-4x-8\over x^2}
\ln(1+x) +\right. \nonumber \\
&& \left.+{1\over 2} + {8\over x} -{1\over 2(1+x)^2}\right] ,
\end{eqnarray}
where:  $x = 2 p^\nu k_\nu$, $p^\nu$ is the electron four-momentum,
and $k_\nu$ is the photon four-wave vector.  The scattering rate (in
the ``lab frame" $K$) by an electron travelling with velocity
$\vec\beta$ in a photon field is given by
\begin{equation}
{\cal R} = c \int d\Omega \int \varepsilon^2 d\varepsilon \,\sigma_{KN}
f_\varepsilon (1 - \vec\beta\cdot\hat n_i)
\end{equation}
where $\Omega$ is the solid angle subtended by the source of soft
photons, $\sigma_{KN}$ is the total cross section,
$f_\varepsilon$ is the photon occupation density, and $\hat n_i$
is the unit vector ($\hat n_i \equiv \vec k/|\vec k|$)
representing the direction of the incoming soft photon.  Taking
the Planck function for $\varepsilon^2f_\varepsilon$ this
equation reads:
\begin{equation}
{\cal R} =  {c\over 4\pi^3} \left({m_e c\over \hbar}\right)^3 
\int d\Omega \int d\tilde\varepsilon \,\sigma_{KN} (1-\vec\beta\cdot\hat n_i)
{\tilde\varepsilon ^2 \over \exp{\tilde\varepsilon/\tilde T} -1 }
\label{rate}
\end{equation}
where $\tilde\varepsilon$ is the photon energy in the
units of electron mass, $\tilde T $ is the photon temperature in
the units of electron mass. 

We draw the energy and direction of the incoming photon from the
distribution of $\varepsilon^2f_\varepsilon$, and $\Omega$,
respectively (as used in eq.6).  Then we transfer them to the
electron instantaneous rest frame $K_{\rm e}$ and there we find
the photon scattering angle $\theta$ using the Klein-Nishina
differential cross section
\begin{eqnarray}
{ d\sigma_{KN}\over d\mu d\phi} &=& {3\sigma_T\over 16\pi}
{1+\mu^2 \over \left[1+{x\over 2}(1-\mu)\right]^2}\times \nonumber \\
&&\times{\left\{ 1+ {x^2(1-\mu)^2\over  4(1+\mu^2)\left[1+{x\over
2}(1-\mu)\right]  } \right\}},
\end{eqnarray}
where $\mu$ is the cosine of $\theta$.
The energy $\epsilon'$ of the scattered photon
in $K_{\rm e}$ is obtained from the Compton formula
\begin{equation}
\epsilon' = {\epsilon \over 1 + {\displaystyle\epsilon\over m_e c^2} (1 -\mu)}.
\end{equation}
Finally, the energy of the outgoing photon is transformed 
back to $K$.

At each step, as the location of the electron changes, the solid
angle $\Omega$ is corrected for the finite size of the source of
soft photons (the polar cap or the stellar surface).  Thus, we
take into account the effect of the decreasing photon number
density with height (geometrical dilution) and then we calculate the
scattering rate ${\cal R}$ at a given height above the neutron
star surface.

\subsection{Magnetic photon absorption}

Despite low magnetic field values, two conditions for the process
of pair creation via magnetic photon absorption ($\gamma + \vec B
\rightarrow e^\pm$) - the energy threshold condition, and the
high optical thickness $\tau_{\gamma B}$ of the magnetosphere -
may well be met in the context of millisecond pulsars.  For
instance, in the model spectra calculated numerically by Rudak \&
Dyks (1999) photons with the energy $\sim 10^2\GeV$ or higher
were subject to the magnetic-absorption reprocessing.

In our simulations we use the absorption coefficient as
described by Erber (1966):
\begin{equation}
\eta(\varepsilon) = {1\over 2} {\alpha \over \lambdabar_{\rm c}} {B_\perp
\over B_{crit}} T\left(\chi \right)
\label{eta}
\end {equation}
where $\alpha$ is a fine structure constant, $\lambdabar_{\rm c}$
is a Compton wavelength, $B_\perp$ is the component of the
magnetic field perpendicular to the photon momentum, and $\chi
\equiv {1\over 2}{B_\perp \over B_{crit} }{\varepsilon\over
m_e c^2}$ is the Erber parameter $\chi$.  For the function $T(\chi)$ we use
its approximation $T(\chi) \approx 0.46 \exp\left(-4 f/3\chi
\right)$, 
 valid for $\chi \simless 0.2$ (for $\chi \simgreat 0.2$
this approximation starts to overestimate $\eta$).
The function $f$ is the near-threshold correction  introduced
by \cite{1983ApJ...273..761D}. In the case of millisecond pulsars
equation~(\ref{eta}) is a good approximation 
even with $f = 1$, since magnetic photon absorption occurs well
above the threshold.
Electron-positron pairs created through the magnetic absorption
radiate then via the synchrotron process.

For demonstrative purposes it is useful to present the condition
$\tau_{\gamma B} \ge 1$ in an analytical way.  It is
straightforward to show that a photon created with a momentum
parallel to the local magnetic field line at a height $h$ above
the neutron star surface will undergo magnetic absorption if its
energy satisfies approximately the following inequality:
\begin{eqnarray}
E & \ge & 10^5 \left({P\over 10^{-3}{\rm s}}\right)^{1/2} 
\left( {B_{pc}\over 10^9{\rm G}}\right)^{-1}
\left({R_{ns}\over 10^6{\rm cm}}\right)^{-1/2} \times \nonumber\\
& &\times \left( 1 + {h\over R_{ns}}\right)^{5/2}\, {\rm MeV.}
\label{tauone}
\end{eqnarray}
This formula is valid for dipolar magnetic field, and the field
line is attached to the outer rim of the polar cap, and it is 
in very good agreement with numerical calculations of magnetic photon absorption.

\begin{figure*}
{\centering \leavevmode
\epsfxsize=.45\textwidth \epsfbox{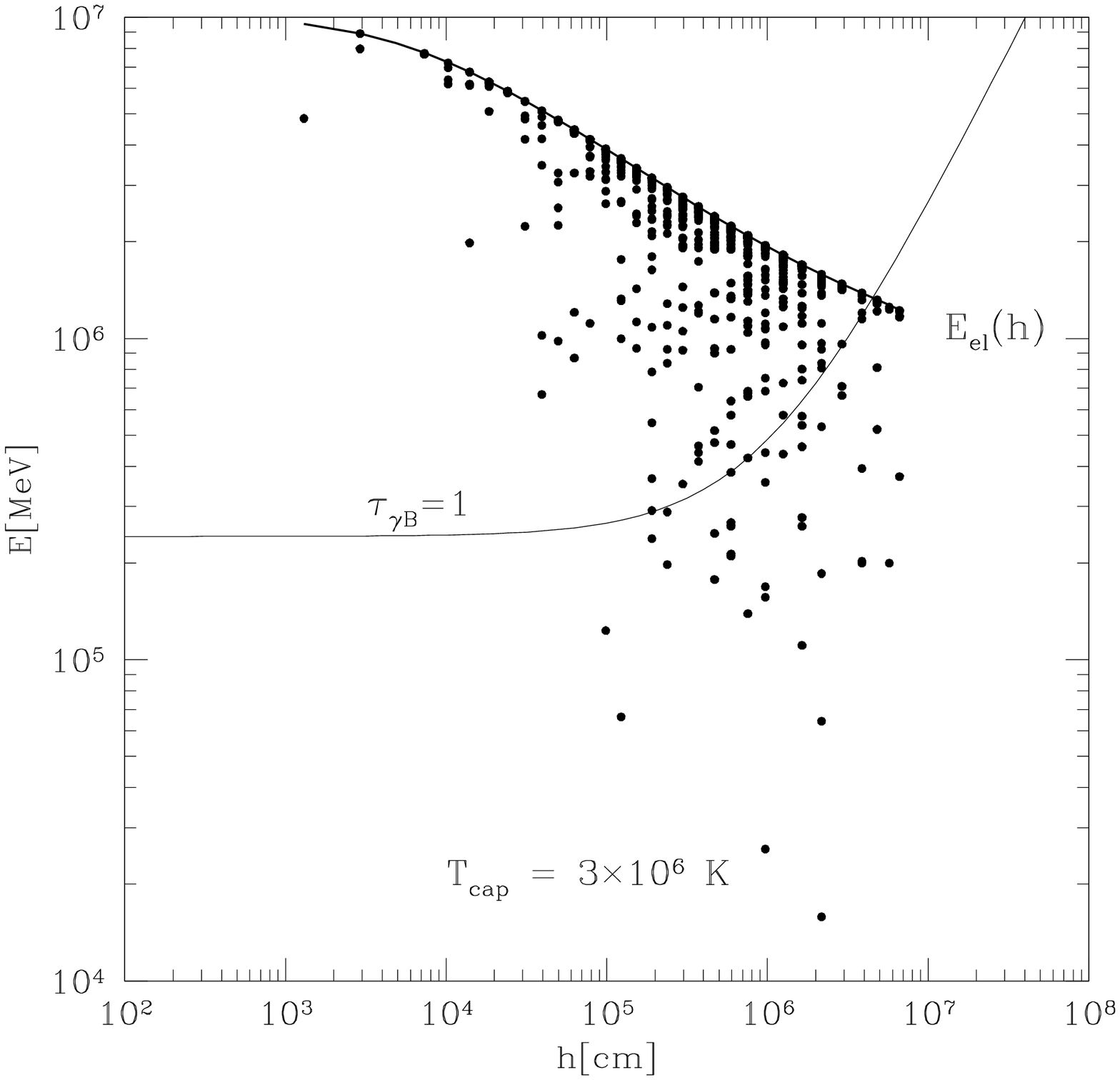} \hfil
\epsfxsize=.45\textwidth \epsfbox{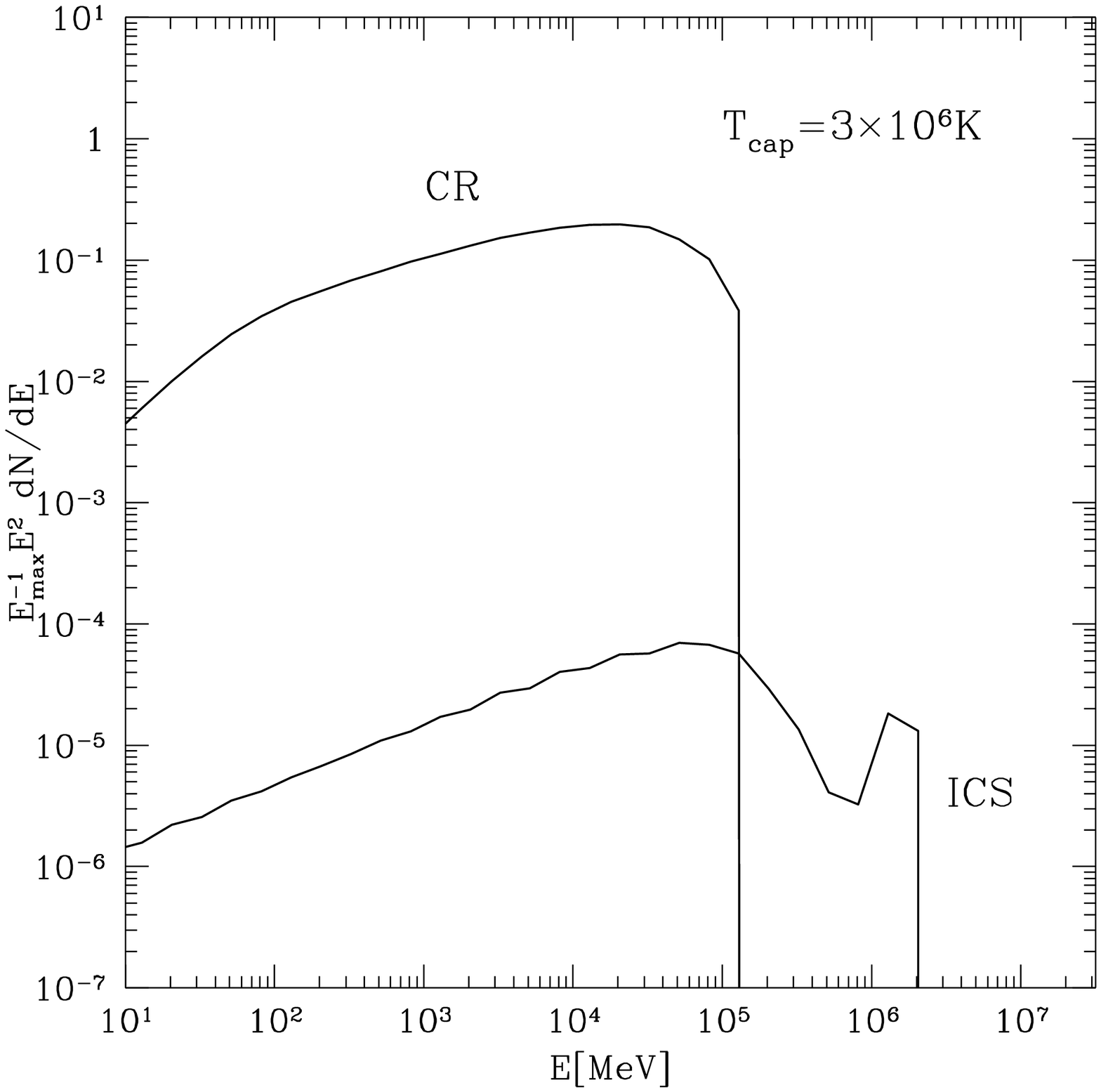}}
{\centering \leavevmode
\epsfxsize=.45\textwidth \epsfbox{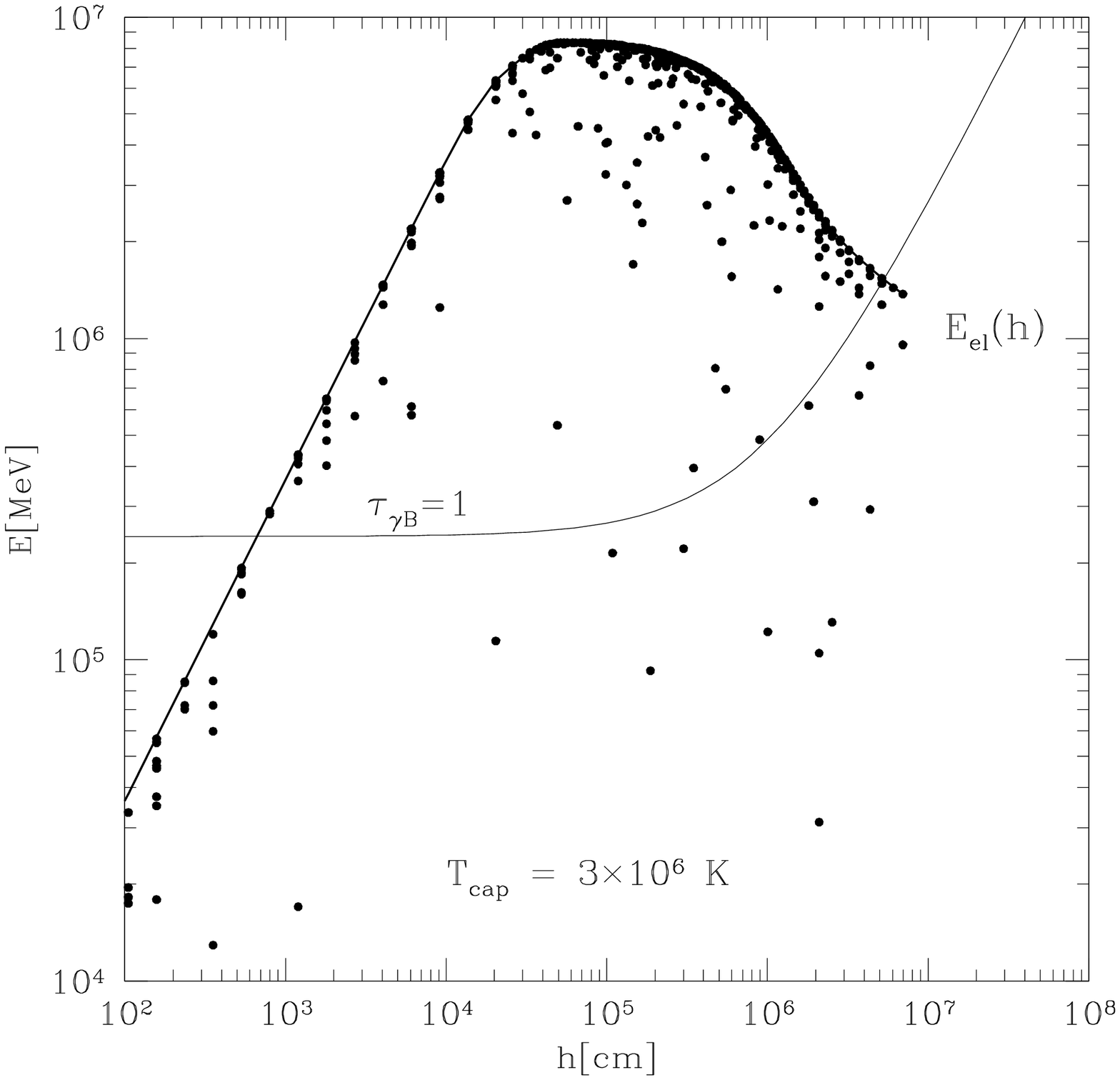} \hfil
\epsfxsize=.45\textwidth \epsfbox{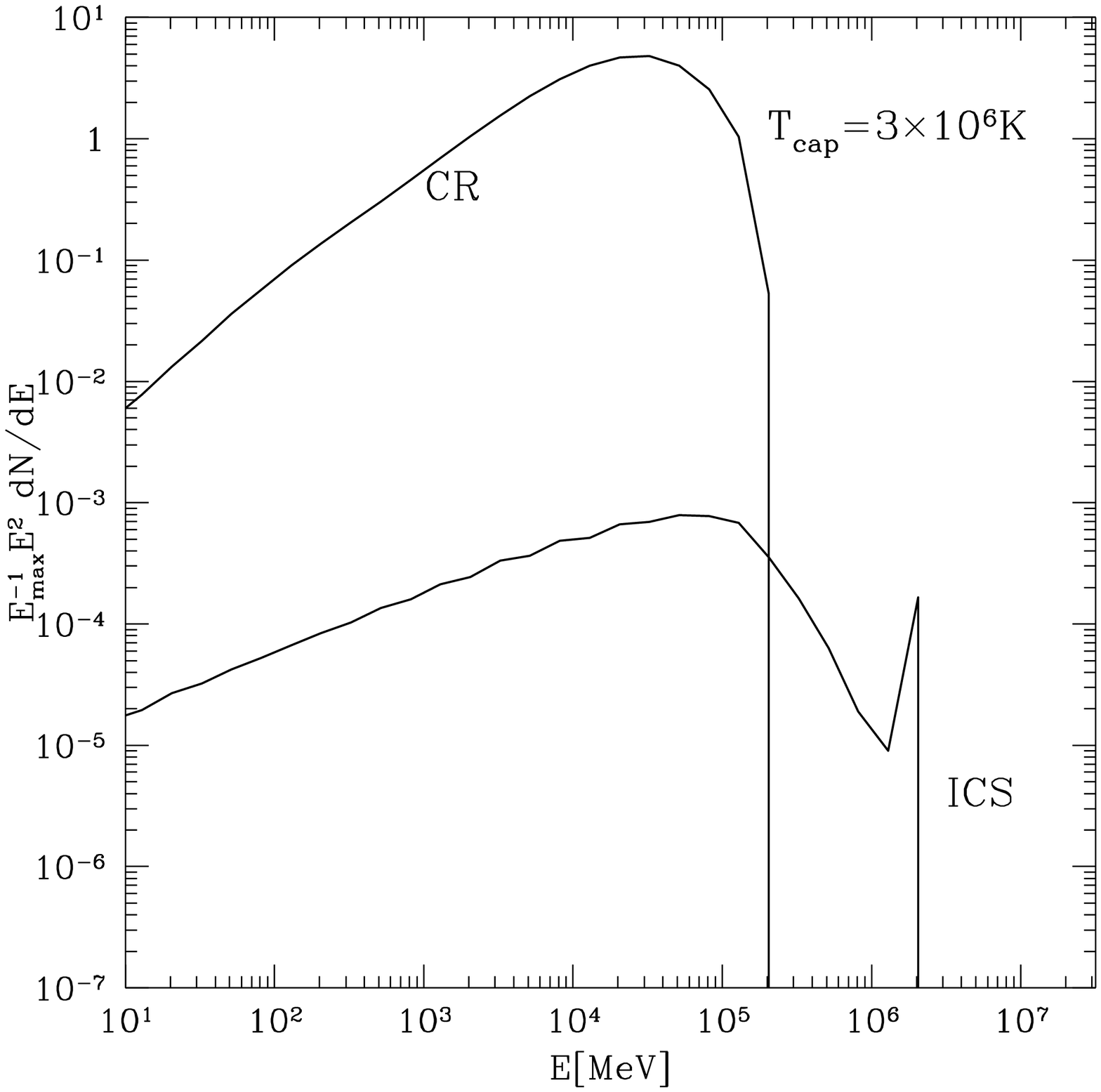}}
\caption{Same as Figure~\ref{fig3} but for the case of thermal
emission from the polar cap with the temperature $3\times 10^6\K.$}
\label{fig4}
\end{figure*}

\section{Results}

Electrons are either accelerated instantly (model A) or
accelerated by the electric field given by equation~\ref{efield}
(model B), and decelerated mainly by the curvature radiation.  In
model B the deceleration rate is insignificant near the polar cap
because it increases with the electron energy and $E_{e}^{\rm
init}$ is small.  On the other hand, the acceleration rate does
not depend on the electron energy and it ceases at the height $h
\sim r_{\rm pc}$ (eq.\ref{efield}).  There exists a region in
between where both competing processes balance.  An example
calculation of the electron energy is shown in Figures~\ref{fig3}
and~\ref{fig4}.  The top left panel of each figure presents
electron energy within model A, with $E_{e}^{\rm init}= 1.07 \times
10^7\MeV$.  The bottom left panel shows the electron energy
within model B; the electron reaches a maximum energy of a few
TeV at the height $h \simeq 4\times 10^4\cm$ (i.e.  $h/r_{\rm
pc}\simeq 1/6$).

The optical depth $\tau_e$ for nonmagnetic scattering of an
electron is small (Figure~\ref{fig1}) and thus scattering is not
a relevant decelaration process of electrons.  The optical depth
for scattering scales as $\propto T^2$, where $T$ is the
temperature of the radiation field.  There are several factors
that influence the scattering rate as a function of the height
above the surface of the neutron star $h$.  The photon density
decreases with height, but the electron spends much more time in
the upper part of the magnetosphere, thus increasing the chance
of scattering there.  
 We find that the latter effect is dominant
and most of the scatterings take place in the upper part of the
magnetosphere between $10^5$ and $10^7\,$cm, where the electron
energy is close to the value of a few TeV.  This is
illustrated in Figures~\ref{fig3} and~\ref{fig4} (left panels),
where the electron energy is plotted as a function of $h$. 
We show the energies of the ICS photons and the locations of the
scattering events.  The intrinsic spectrum of the ICS photons 
has a width of about one decade in energy centered around a few TeV, 
corresponding to the
range of electron energies in the region where most of the
scattering occur.
In Figure~\ref{fig2} we illustrate the effects of reprocessing in the pulsar
magnetosphere (taking model A as an example) and present both,
the intrinsic spectrum of ICS photons (dotted line) and the spectrum after reprocessing
due to magnetic absorption (solid line). Magnetic
absorption removes the high energy part of the intrinsic ICS spectral peak,
decreasing the strength of the peak by a factor of more than 10,
and redistributing the ICS photon energy to lower part of the
spectrum.  The ICS photons subject to magnetic absorption
reprocessing are located in Figures~\ref{fig3} and~\ref{fig4}
(left panels) above the thin line corresponding to
$\tau_{{\gamma}B}=1$ (equation~[\ref{tauone}]).  In the case when
the soft photon source is the hot polar cap most of the ICS
photons will be reprocessed; see Figure~\ref{fig4}.  However, in
the case when the photon source is a warm surface of a neutron
star most of the scattering events take place at a distance of a
few neutron star radii from the surface and most of the ICS
photons will escape freely; see Figure~\ref{fig3}.  The escaping
ICS photons have typical energies equal to the electron energy at
the height of few neutron star radii above the stellar surface
and thus the spectrum of the ICS photons exhibits a strong peak
at the energy of $\approx 1\,$TeV.  The height of the ICS peak
depends on the number of the soft thermal photons in the
magnetosphere, and so it is directly related to the temperature
and the radiating area (polar cap or entire stellar surface).

Another factor that influences the scattering rate is the soft
photon flux  anisotropy. Such anisotropy results from opacity
angular dependence  in strong magnetic field, and cannot be neglected in the
case of classical pulsars with $B\approx 10^{12}\,$Gauss.
Detailed models of magnetized atmospheres show that in the case $10^9\,$Gauss
fields the soft photon  flux
is almost constant up to  about 60 degrees \cite{1994A&A...289..837P}.
This result justifies our simplified treatment of the soft photon field
as being isotropic.

The dominant spectral component in the energy range from
$10\MeV$ out to its cutoff at $\simgreat 100\GeV$ is due to
curvature radiation (see the energy spectra in
Figures~\ref{fig3} and~\ref{fig4}, right panels). This cutoff
energy is significantly higher than that expected in some
classical pulsars as well as measured in some of them ($\approx
10\,$GeV, Thompson et~al. \shortcite{1997comp.symp...39T}). 
This is because of smaller curvature radii implying higher
values of the characteristic photon energy, and also much weaker
magnetic absorption ($\gamma B \rightarrow e^\pm$). The slope
$\alpha$ of radiation energy spectrum per logarithmic energy
bandwidth ($E^2 dN/dE \propto E ^{\alpha}$) below $\epsilon_{\rm
break} \approx 100\MeV$ is insensitive to model of acceleration,
and is close to $4/3$ (see Rudak \& Dyks 1999 for discussion of
$\epsilon_{\rm break}$).  Above $\sim 100 \MeV$ the spectrum
assumes a slope which depends on the acceleration model. For
model A the slope between $100\MeV$ and $10\GeV$ is $1/3$, in
agreement with simple analytical estimates for monoenergetic
source function of beam particles. The slope for model B was
found numerically to be equal $0.84$ \cite{1999..BuRu}.  The
electron  acceleration in model B makes the electron spectrum
harder in comparison to the case of model A, which in
consequence leads to a harder CR spectrum. Note, that the total
energy contained in gamma rays is larger in model B than in
model A (compare  right-hand panels of either Figure~\ref{fig3}
or  Figure~\ref{fig4}). In model A the initial energy of
electrons is large enough to enable pair creation, while in
model B they need to overcome the CR losses to attain
$E_{e}^{\rm max}$.  The work done by the electric field on the
electron in model B is about 20 times larger than $E_{e}^{\rm
max}$.

\begin{figure}
{\centering \leavevmode
\epsfxsize=\columnwidth 
\epsfbox{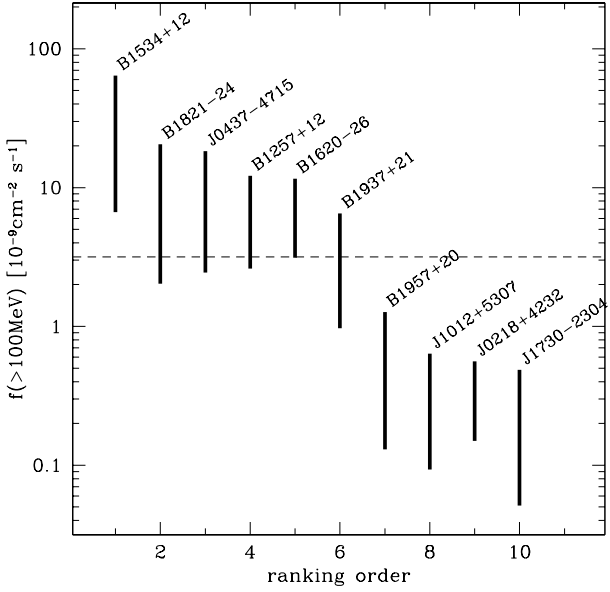}}
\caption {Millisecond 
pulsars ranked according to their gamma ray fluxes
above $100$\,MeV expected within model A. Each vertical bar spans the range in the
flux when the energy of injected primary electrons
$E_{e}^{\rm
init}$ varies 
by a factor of two. The horizontal dashed line represents 
the \glast sensitivity.
}
\label{fig6}
\end{figure}

\begin{figure}
{\centering \leavevmode
\epsfxsize=\columnwidth 
\epsfbox{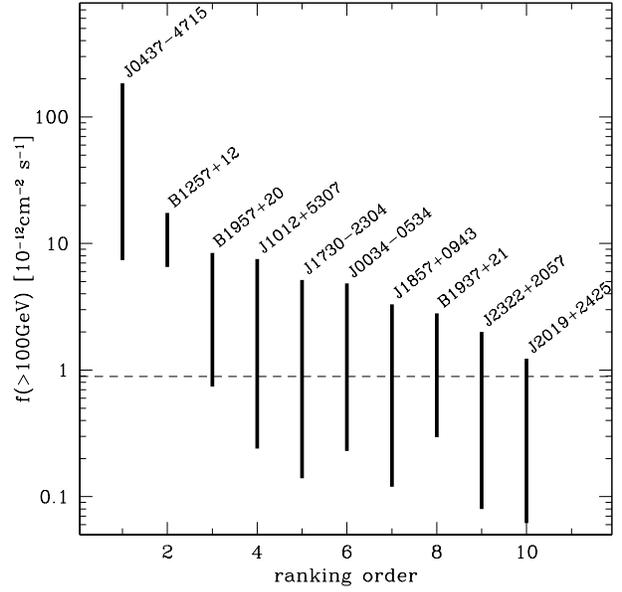}}
\caption {Similar to Fig.6 but the ranking is for gamma ray fluxes
above $100$\,GeV, and the horizontal dashed line represents 
the \magic sensitivity.
}
\label{fig7}
\end{figure}

\section{Discussion}

We calculated energy spectra of pulsed HE and VHE emission from
millisecond pulsars for two models of beam particles
acceleration.  The spectra are a superposition of curvature (CR),
synchrotron (SR), and Compton upscattering (ICS) components.  The
curvature component dominates the region between 1\,MeV and
100\,GeV.  The slope of the spectrum in the range between
$100\,$MeV and $10\,$GeV is sensitive to the details of the
electron acceleration process.  The synchrotron component becomes
important only below $\sim 1\MeV$ (Rudak \& Dyks 1999).  The ICS
component has a narrow peak around 1\,TeV and its strength is
related to the number density of the soft photons in the
magnetosphere.  Two sources of soft photons were considered:  the
hot polar cap with a uniform temperature $T_{\rm pc}\simgreat
10^6\K$, and the entire surface of the neutron star, with a lower
temperature:  $T_{\rm surf}\simgreat 10^5\K$.

\begin{figure}
{\centering \leavevmode
\epsfxsize=\columnwidth 
\epsfbox{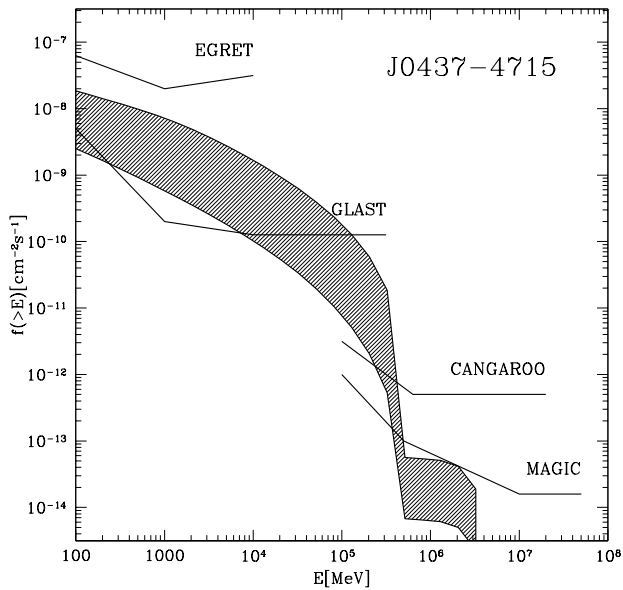}}
\caption {
Integral spectral fluxes of photons calculated for J0437-4715 at the distance of $140\,$pc. 
Model A is used with two following values of $E_{e}^{\rm
init}$: 1) $1.54\times 10^7\,$MeV (after Rudak \& Dyks 1998, see also section 2.2),
2) a value two times higher, i.e. $3.08\times 10^7\,$MeV.  
These two values give the spectra (solid lines) which limit the shaded region 
from the bottom and the top, respectively.
The soft photons used for ICS
come from the
entire stellar surface with the temperature $4\times 10^5\K$
\protect\cite{1995ApJ...454..442E}.
Sensitivities of several high energy 
detectors are also indicated.}
\label{fig5}
\end{figure}

Several new gamma-ray telescopes will be operational in the near
future \cite{1999astro-phWee} and thus it is interesting to
discuss the observability of the HE radiation from millisecond
pulsars. For our ranking we have selected the millisecond
pulsars from the pulsar catalog of Taylor et al. (1995)
\nocite{princeton}  using the following criteria $P< 60\,$ms and
$\dot P< 10^{-17}$. For each pulsar  the expected integral
fluxes above $100\,$MeV and above $100\,$GeV was then
calculated.  Model A had been chosen for these calculations as a
more conservative one. Moreover,  two cases were considered for
each pulsar: in addition to the case with the energy of injected
primary electrons $E_{e}^{\rm init}$ being chosen according to
section 2.2 (see also eq.8 of Rudak \& Dyks 1998),  we also
allowed  the electron energy to be two times higher.  This was
done to present how sensitive the results are  to our choice of
the electron's energy. We used the results of Rudak \& Dyks
\shortcite{1998MNRAS.295..337R} to callibrate the fluxes of
primary electrons (note that the spectra presented in
Figures~\ref{fig3} and~\ref{fig4}, are per single electron). 
Ten best candidates in each band are presented in
Figures~\ref{fig6} and~\ref{fig7} respectively. The first two
candidates for GLAST observations (Figure~\ref{fig6}), B1534+12
and B1821-24,  are objects with relatively  high magnetic
fields, which place them in a transition region between the
domains of classical and millisecond pulsars. Because of the
magnetic field strength and consequently high magnetic
absorption in these two objects, the high energy cutoff lies
below  $100\,$GeV and therefore they disappear from the ranking
of Figure~\ref{fig7}.

 The third best candidate for GLAST observations is J0437-4715,
a pulsar located very close to the Earth  at the distance of
$140\,$pc, with the spin period $P=5.75\,$ms, and the
magnetic field at the polar cap 
$B_{\rm pc} =7.4\times 10^8\,$G inferred from kinematically
corrected $\dot P$ \cite{1994ApJ...421L..15C}.  J0437-4715 is
also our primary candidate for observations in the range above
$100\,$GeV with the future  Cherenkov telescopes.  The integral
photon spectra of J0437-4715 obtained in the course of
calculations for the purposes of the ranking  are presented in
Figure~\ref{fig5} along with the sensitivities of GLAST,
CANGAROO, as well as MAGIC in its `Large Zenith Angle' mode. 
The expected photon flux in the \egret energy range (i.e. 
between $0.1$ and $10\,$GeV) is between $\approx 2.0\times
10^{-9}\,$cm$^{-2}\,$s$^{-1}$, and $\approx 2\times
10^{-8}\,$cm$^{-2}\,$s$^{-1}$  (for $E_{e}^{\rm init}=1.54\times
10^7\,$MeV and  $E_{e}^{\rm init}=3.08\times 10^7\,$MeV,
respectively) and falls below the upper limit of $ 2.1\times
10^{-7}\,$cm$^{-2}\,$s$^{-1}$ placed with \egret for this object
\cite{1995ApJ...447..807F}.  Figure~\ref{fig5} shows that the
flux expected from J0437-4715   in the   \glast energy range
($0.1$ to $200\,$GeV) is above its sensitivity limit of $3\times
10^{-9}$cm$^{-2}\,$s$^{-1}$ \cite{1999astro-ph}.

In the list of  potential targets for atmospheric Cherenkov
telescopes,  presented in Figure~\ref{fig7}, J0437-4715 clearly
stands out as the strongest source, mainly because of its
proximity. The next  source B1257+12 is also rather strong and
should be detected  by the telescopes in the northern hemisphere
like MAGIC and VERITAS, and also it could be seen by CANGAROO-III.
The remaining objects have the expected fluxes between a few
times $10^{-13}$ and a few times $10^{-12}$cm$^{-2}$s$^{-1}$,
which places them in a class of possible but uncertain targets.

The 100 GeV part of J0437-4715 spectrum can be probed by
Cherenkov telescopes in the southern hemisphere like
CANGAROO-III.  This object should be also visible just above the
horizon by the \magic telescope in the Canary Islands. The
expected photon flux above $100\,$GeV is $\approx
10^{-10}$cm$^{-2}\,$s$^{-1}$ which is almost two orders of
magnitude above the sensitivity of \magic in its Large Zenith
Angle mode \cite{1998..Magic}. \magic may also probe the ICS
component of the spectrum (above $0.5\,$TeV) provided the
energy  of beam particles (i.e. primary electrons) is higher
than assumed in our model.

Our ranking list of Figure~\ref{fig6}  also includes J0218+4232,
a pulsar with a likely detection by \egret at about $3\sigma$
level  (Verbunt et al. 1996, Kuiper \& Hermsen
2000)\nocite{agn-pulsar}.  Our theoretical gamma ray flux for
this object is a factor of up to ten below the expected \glast
sensitivity, mainly because of the large distance. Thus within
the model presented above we remain skeptical to this report
about its positive detection, knowing that the gamma ray source
is coincident with an AGN.

Millisecond pulsars are potentially very interesting sources of
high energy gamma-rays.  The gamma-ray radiation, if detected,
shall provide useful information about physical conditions in a
pulsar magnetosphere:  the primary electron energy, the nature of
the electric field accelerating the primary electrons, and
indirectly soft photon density.  We appeal to include the
millisecond pulsars into target lists for the upcoming gamma-ray
observatories, both space- and ground-based.

\section*{Acknowledgments}

This reasearch has been supported by KBN grant 2P03D02117.
We thank Gottfried Kannbach for useful discussions on gamma-ray
detectors.
Insightful comments and suggestions made by an anonymous referee
allowed us to improve considerably the presentation of the paper.

\newcommand{\apj}{ApJ}
\newcommand{\apjl}{ApJ}
\newcommand{\mnras}{MNRAS}
\newcommand{\aap}{A\&A}

\end{document}